\documentclass[runningheads]{llncs}
\usepackage{amsmath,amssymb,amsfonts,stmaryrd,extarrows,dsfont}
\usepackage{graphicx}
\usepackage{subfigure}
\usepackage{cite}

\begin{document}
\title{
Collaborative Control Method of Transit Signal Priority Based on Cooperative Game and Reinforcement Learning
}

\author{Hao Qin \and
Weishi Zhang\thanks{teesiv@dlmu.edu.cn}
}

\institute{Information Science and Technology College, Dalian Maritime University, Dalian, Liaoning, China, 116026}

\maketitle              %
\begin{abstract}
To address the low efficiency in priority signal control within intelligent transportation systems, this study introduces a novel eight-phase priority signal control method, CBQL-TSP, leveraging a hybrid decision-making framework that integrates cooperative game theory and reinforcement learning. This approach conceptualizes the allocation of bus signal priorities as a multi-objective decision-making problem across an eight-phase signal sequence, differentiating between priority and non-priority phases. It employs a cooperative game model to facilitate this differentiation. The developed hybrid decision-making algorithm, CBQL, effectively tackles the multi-objective decision-making challenges inherent in the eight-phase signal sequence. By computing the Shapley value function, it quantifies the marginal contributions of each participant, which in turn inform the construction of a state transition probability equation based on Shapley value ratios. Compared to conventional control methods, the CBQL-TSP method not only upholds the fairness principles of cooperative game theory but also harnesses the adaptive learning capabilities of Q-Learning. This enables dynamic adjustments to signal timing in response to real-time traffic conditions, significantly enhancing the flexibility and efficiency of priority signal control.

\end{abstract}

\section{Introduction}\label{sec1}

Traffic Signal Priority (TSP) control methodologies represent a critical domain of investigation within the nexus of traffic engineering and Bus Rapid Transit (BRT) systems \cite{bib1}. These methodologies are paramount in their pursuit of augmenting the efficiency and safety of transportation systems, particularly with a focus on optimizing the flow of public transport and emergency vehicles. Prevailing paradigms in traffic signal control, predicated on fixed-time signal schedules, exhibit limitations in adaptability and efficacy \cite{bib2}. These schedules manifest inefficiencies, inducing unwarranted delays during periods of subdued traffic demand and exacerbating congestion during peak demand.

This paper introduces an innovative eight-phase priority signal control paradigm expressly crafted for BRT systems, with the overarching goal of concurrently optimizing private vehicle stability and the reliability of bus services. Departing from conventional methodologies, our proposed approach intricately accounts for bus signal priority, strategically engineered to confer maximal stability upon private vehicular traffic while safeguarding the dependability of bus services.

At the crux of our investigation lies the fusion of cooperative gaming principles with reinforcement learning mechanisms, orchestrating collaborative decision-making processes bus and private vehicle prioritization. This amalgamated framework endows our system with the dynamism required for real-time decision adjustments, thereby fostering an adaptive and responsive traffic signal control infrastructure. The decision-making apparatus encompasses predetermined conditions and a discretized bus strategy, ensuring contextualized prioritization informed by local conditions at each intersection.

To empirically scrutinize the efficacy of our proposed approach, we conducted simulations on a veritable road network featuring a BRT system. Through meticulous simulation analyses, we assess the nuanced impact of our eight-phase priority signal control paradigm on traffic dynamics, stability metrics, and the reliability of bus services. This research not only introduces a pioneering facet to the landscape of TSP control but also furnishes pragmatic insights germane to its integration within veritable computer science-infused transportation networks.

\section{Related Work}\label{sec2}

Prioritizing public transport in urban planning enhances construction and management to meet growing demand and address issues like congestion. This focus not only supports greater capacity compared to private vehicles but also conserves resources and improves the urban environment. Additionally, TSP Control is essential in traffic engineering, particularly for optimizing public and emergency vehicle efficiency through intelligent signal control at intersections, crucial for reducing traffic congestion. Intelligent bus priority signal control at intersections is crucial for mitigating traffic congestion and comes in three main types.
\subsection{Passive priority control}\label{sec2.1}

The concept of passive priority, introduced in 1973 by Sperry Rand Corporation, involves adapting intersection signals to historical bus route patterns without real-time vehicle tracking, making it cost-effective as it requires no special detection equipment \cite{bib3}. However, lacking real-time data can reduce the effectiveness of passive priority unless there's a robust understanding of public transport flows and schedules \cite{bib4}. This method is mainly applied at intersections with frequent bus traffic, using established passive priority logic to manage frequent priority requests \cite{bib5, bib6, bib7}.

\subsection{Proactive priority control}\label{sec2.2}
Proactive Priority Control is the installation of sensing devices on buses or at intersections to improve traffic state recognition. The cost of proactive priority control is higher compared to passive methods, but its effectiveness in prioritizing public transit, coupled with lower societal and network-level impact, justifies its exploration.Current research delves into the complexity of trigger conditions for proactive priority strategies. The classification of trigger conditions includes distinctions between manual \cite{bib8,bib9} and automatic control, with the latter further categorized into rule-based \cite{bib10} and model-based decision triggers \cite{bib11,bib12,bib13}. The goal is to achieve adaptive and responsive TSP control systems that consider real-time conditions and improve the overall efficiency of transportation networks.In examining signal adjustment strategies, existing literature highlights two prominent proactive approaches: "red signal truncation or green signal advancement" \cite{bib14} and "green signal extension" \cite{bib15}. The former strategically shortens signal phase durations when buses approach intersections with red signals, reducing waiting times for buses in the red signal direction. The latter involves extending green signal times for impending signal changes, ensuring the smooth passage of buses through intersections. Researchers have also investigated complex control strategies, while demanding stronger traffic perception and signal adjustment capabilities, pose challenges due to their potentially more aggressive nature in phase switching \cite{bib16,bib17}. Despite the potential safety risks, these strategies contribute to the ongoing discourse on innovative TSP control methods.

\subsection{Adaptive priority control}\label{sec2.3}
Adaptive priority control emphasizes the immediate responsiveness of the system to the current state of the traffic environment. In this approach, adjustments to signal timings and priority decisions are made promptly based on real-time conditions. It represents a nuanced aspect of proactive control, underlining the system's quick adaptability \cite{bib24}. Real-time priority control is the use of road detection, vehicle positioning technology, etc., to obtain the region's road traffic conditions and the operating status of public transport vehicles, so as to carry out real-time control of signals to ensure that public transport priority. Road traffic conditions, including the average speed of vehicles in the area, traffic conditions at various intersections, and bus operations \cite{bib18}.Swinburne University of Technology, Australia, and Monash University jointly proposed a mathematical framework to establish a TSP system based on a single-objective network and a dynamic traffic assignment method \cite{bib19}. It mainly investigates the priority right-of-way problem in the case of mixing of priority and non-preferred vehicles.The University of Minnesota proposed a maximum pressure signal control method to implement adaptive TSP control for BRT to provide priority constraints for buses in order to achieve maximum stability and reliable bus service for nonpreferred vehicles \cite{bib20}.Yang et al proposed a TSP algorithm for multimodal traffic control using interconnected vehicle information \cite{bib21}. Paul Anderson and Carlos F. Daganzo introduce Conditional Signal Priority (CSP) as a targeted TSP approach to address bus bunching. Through mathematical modeling and simulations, CSP is shown to notably enhance bus reliability, reduce traffic delays, and outperform traditional TSP methods in high-frequency systems \cite{bib22}.

In the ensuing investigation, endeavors will be directed towards ameliorating the challenges identified, with a specific focus on enhancing the adaptability, optimizing priority allocation, and improving the efficiency of traffic signal priority systems, particularly within the context of mixed traffic conditions.

\section{Design of Eight-phase signal control method}\label{sec3}

\subsection{Problem Description}\label{sec3.1}

This paper introduces a novel TSP control method, CBQL-TSP (Cooperative Bargaining games and Q-Learning for Transit Signal Priority), which models the traffic signal problem of TSP systems as a multi-agent, multi-action space Markov model. The approach utilizes cooperative bargaining strategies to address the challenges related to priority signal and non-signal decisions within the model. The optimization objectives of the model are to minimize the total delay time and to minimize the maximum delay time of priority vehicles.

\begin{figure}[!t]
	\centering
	\includegraphics[width=8cm]{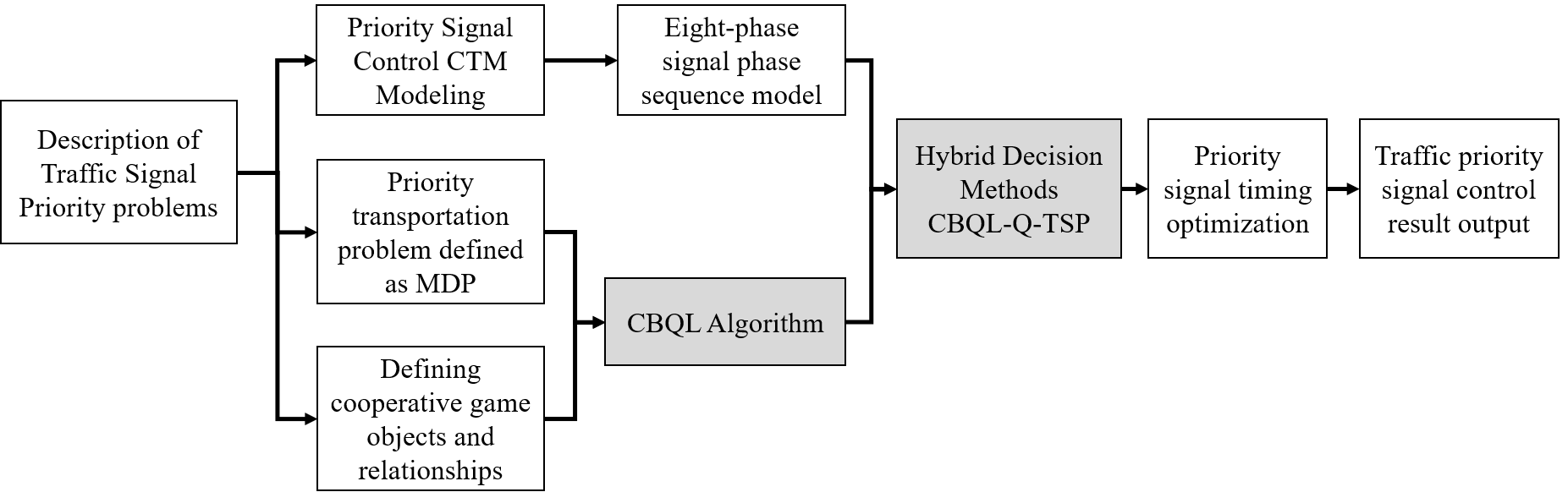}
	\caption{Design process of CBQL-TSP collaborative control method}
	\label{fig:fig1}
\end{figure}

The design process of the CBQL-TSP control method is illustrated in Fig.~\ref{fig:fig1}. Initially, a comprehensive description of the transit signal priority issue is provided. The method incorporates game-theoretic analysis, focusing specifically on the interactions between priority and non-priority vehicles. This analysis lays the groundwork for the implementation of the CBQL algorithm, a dynamic learning approach that adjusts the signal control strategies for both vehicle types to achieve a balance of contributions among multiple agents, ensuring equilibrium and optimal solutions within a game-theoretic framework. Given the complexities of priority signal control, the traffic signal priority environment is configured, and priority signal timings are adjusted based on the optimal strategies derived from the algorithm. The system outputs the results of the transit signal priority control, validating the effectiveness of the CBQL-TSP method. To enhance the system's adaptability, an eight-phase priority signal control method is proposed, and core parameters are rigorously defined and integrated to ensure stability and efficiency in the dynamic control of traffic signals.

\subsection{Traffic Flow Model}\label{sec3.2}	
\begin{figure}[!t]
	\centering
	\includegraphics[scale=0.4]{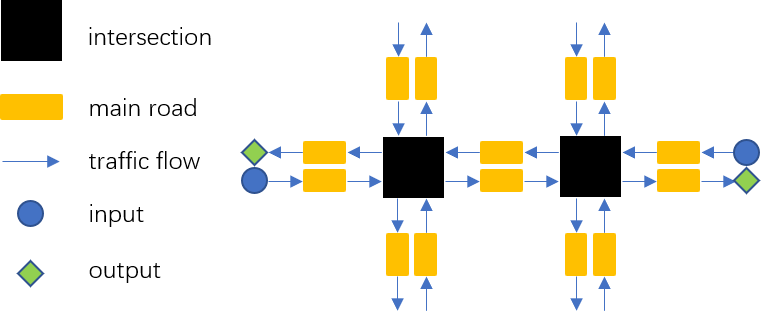}
	\caption{CTM of two intersections}
	\label{fig:fig2}
\end{figure}

The data obtained from short-term traffic flow prediction is used to model the traffic flow based on the commonly used traffic flow model, the Cellular Transfer Model (CTM). For example, in Fig.~\ref{fig:fig2}, a traffic network of 2 intersections containing two origins and destinations is used. The choice of routes between origins and destinations as well as the proportion of traffic flow on each route are incorporated into the CTM. Since the number of intersection phases does not significantly affect the structure of the CTM, the model is easily extendable to asymmetric networks.

\begin{figure}[!t]
	\centering
	\includegraphics[width=8cm]{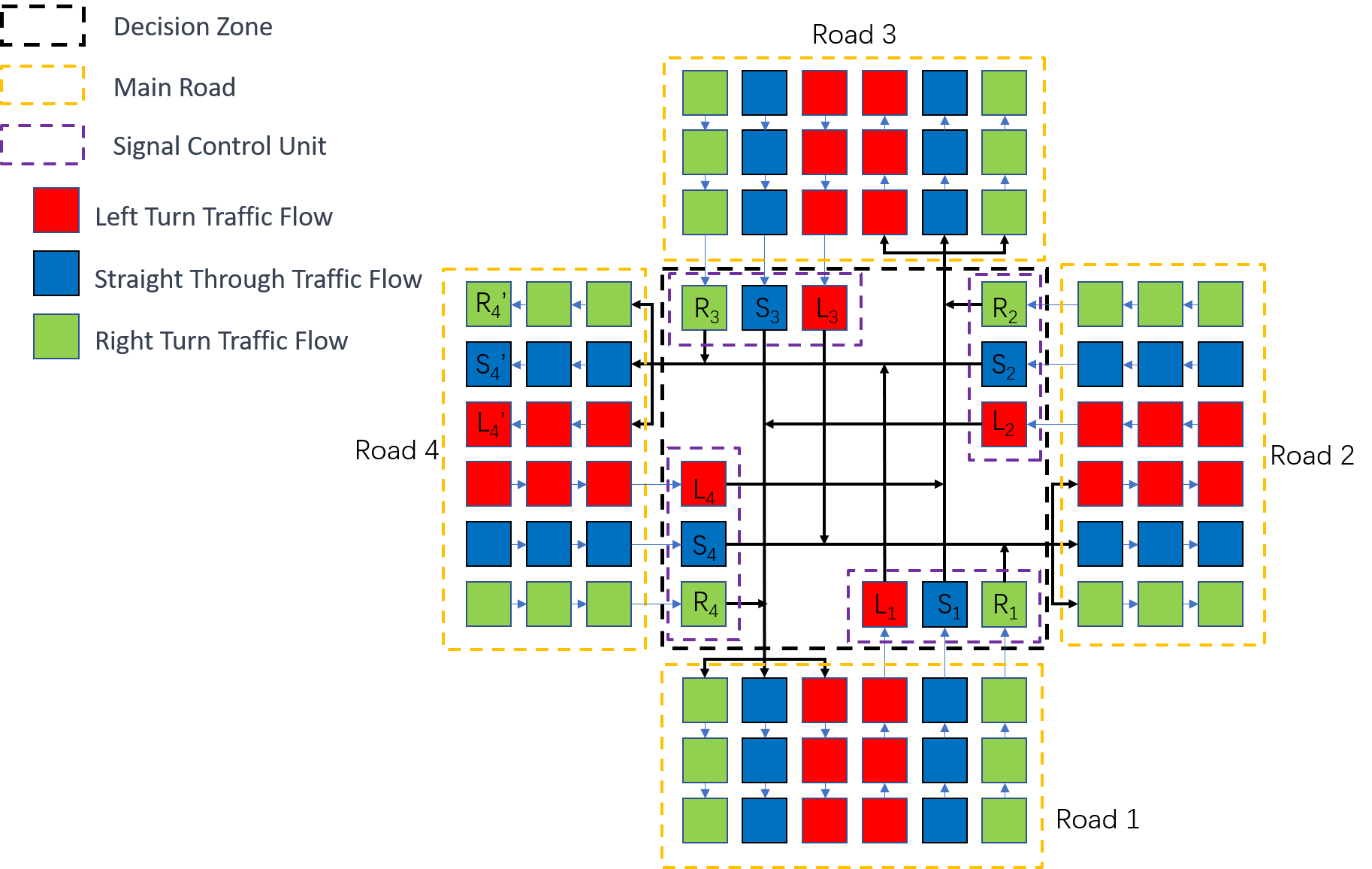}
	\caption{CTM of a single intersection}
	\label{fig:fig3}
\end{figure}

For each intersection, the study uses a cell representation of a link intersection. In the link, the traffic flow is divided into two parts along the forward and reverse directions of movement, the upstream sending area and the downstream receiving area. The traffic flow is divided into left-turn, straight ahead and right-turn traffic flows based on the predicted routes. As shown in Fig.~\ref{fig:fig3}.

In the CTM, roadways are discretized into cells, and the evolution of traffic states occurs over discrete time steps. Within this cellular framework, we'll represent the traffic density, flow, and velocities for buses and non-bus vehicles.

Let $\rho_{bus}^{i}$ and $\rho_{non-bus}^{i}$ represent the densities of buses and non-bus vehicles in cell $i$ at time $t$, and $\rho_{bus}^{i}$ and $\rho_{non-bus}^{i}$ denote the corresponding flow rates. The velocities $v_{bus}^{i}$ and $v_{non-bus}^{i}$ characterize the average speeds of buses and non-bus vehicles in cell $i$.
The CTM-based equations are then expressed as follows:

\begin{align}\label{eq1}
\rho_{\text{bus}}^{(i,t+1)} &= \rho_{\text{bus}}^{(i,t)} + \frac{\Delta t}{\Delta x}(q_{\text{bus}}^{(i-1,t)} - q_{\text{bus}}^{(i,t)}) \\
\rho_{\text{non-bus}}^{(i,t+1)} &= \rho_{\text{non-bus}}^{(i,t)} + \frac{\Delta t}{\Delta x}(q_{\text{non-bus}}^{(i-1,t)} - q_{\text{non-bus}}^{(i,t)})
\end{align}

where:

\begin{align}
q_{\text{bus}}^{(i,t)} &= \min\left(\rho_{\text{bus}}^{(i,t)}, v_{\text{bus}}^{(i,t)} \cdot \Delta x\right) \label{eq2-1} \\
q_{\text{non-bus}}^{(i,t)} &= \min\left(\rho_{\text{non-bus}}^{(i,t)}, v_{\text{non-bus}}^{(i,t)} \cdot \Delta x\right) \label{eq2-2}
\end{align}

The overall density $rho^{i,t}$ in cell $i$ at time $t$ is the sum of bus and non-bus densities:

\begin{equation}
\rho^{(i,t)} = \rho_{\text{bus}}^{(i,t)} + \rho_{\text{non-bus}}^{(i,t)} \label{eq3}
\end{equation}

\subsection{Eight-phase Priority Signal Model}\label{sec3.3}	
\begin{figure}[!t]
	\centering
	\includegraphics[width=4cm]{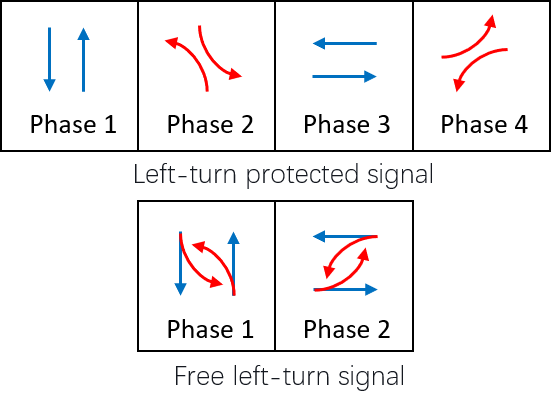}
	\caption{The phase sequence for both types of left-hand signals}
	\label{fig:fig4}
\end{figure}

Two types of left-traffic protection signals and free left-traffic signals are selected for signal phase sequences. As shown in Fig.~\ref{fig:fig4}, when the intersection is highly saturated, a left-turn protection signal is used to protect the left-turning traffic by preventing the opposite direction from going straight. When the saturation is low, the free left signal is used to allow the left-turning traffic to cross the opposite traffic flow by itself.

\begin{figure}[!t]
	\centering
	\subfigure[BusRouteStraight]{\includegraphics[width=0.4\linewidth]{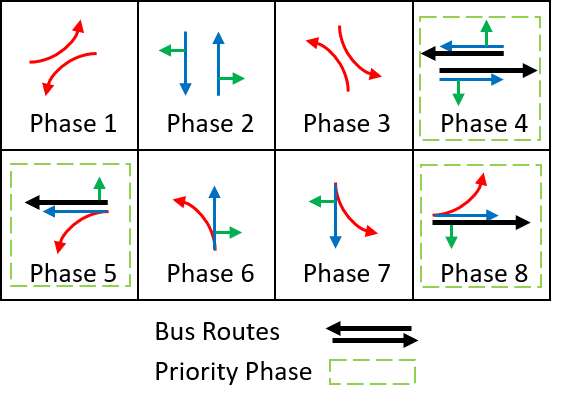}}
	\hspace{1 em} 
	\subfigure[BusRouteLeftTurn]{\includegraphics[width=0.4\linewidth]{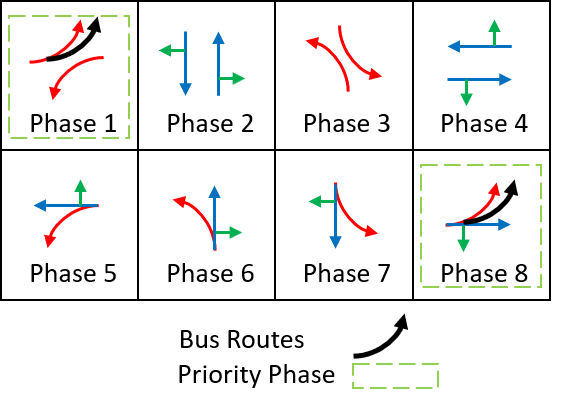}}
	\caption{Phase Sequence Design for Eight-Phase Bus Priority Signals}
	\label{fig:fig5}
\end{figure}

The study attributes the realization of bus signal priority to a multi-step decision-making problem for signal phase sequences. An eight-phase signal control method is proposed to study the signal control phase sequences into priority and nonpreferred signals according to the formulated routes of smart buses. As shown in Fig.~\ref{fig:fig5}, the intersection signals are divided into different eight-phase models according to whether the priority bus route is straight or left, in which phase 1 and phase 8 of the left model are the priority phases containing the priority bus route. Phase 4, 5, and Phase 8 of the straight model are the priority phases containing the priority bus route. Due to the non-sequitur nature of traffic signal timing, the problem is modeled as a Markov decision process (MDP) in the next section.

\section{Research on CBQL-TSP Method}\label{sec4}
\subsection{CBQL algorithm}\label{sec4.1}

In this section, a joint cooperative game and reinforcement learning approach to TSP decision-making is developed, where the interactions between traffic signals in a TSP system can be modeled as a game, where each signal (or group of signals) is a player trying to minimize its delay. Depending on the uncertainty and dynamics of the traffic system, this can be combined with a MDP.

Each agent in the game represents a set of traffic signals (priority or non-priority signals). The state of each agent is defined by the traffic situation, including the presence and urgency of priority vehicles (e.g., buses, emergency vehicles) and the current signal phase. Each participant's action is to request a certain amount of green time. The payoff (or utility) of each participant is a function of its delay, which is intended to be minimized.

To incorporate the dynamics and uncertainty of the transportation system, the study models the TSP system as a two-intelligent body learning model. The model can be defined by a tuple $(S, A, P, R)$:

\textbf{State (S)}: The state of intersection signal control agent \( k \) is represented as a vector \( \mathbf{S}_k[j] \), containing \( P + 3 \) components \( j \), where \( P \) is the number of phases at the current intersection, and \( j = 0, 1, \ldots, P+2 \). The first three components are: the phase currently in the green light stage, the green ratio of the current phase, and the signal cycle duration of the current phase. The remaining \( P \) components correspond to the current traffic flow for each phase. The state vector \( S_k[j] \) is defined as follows:
\begin{equation}\label{eq4}
S_k[j] = 
\begin{cases} 
\psi^k & \text{if } j = 0, \\
\lambda_a^k & \text{if } j = 1, \\
T_a^k & \text{if } j = 2, \\
Q^k[j-3] & \text{if } 3 \leq j \leq P+2.
\end{cases}
\end{equation}

The state space set of the MDP is denoted as \( S = [S_1, \ldots, S_k, \ldots, S_n] \), where \( n \) is the number of intersections. According to the formulae provided, the state space includes the green light phase \( \psi \), the green signal ratio \( \lambda \), the signal cycle \( T \), and the current traffic flow for each phase \( Q[j] \). Therefore, the state vector for intersection \( k \) is given by \( s_k = \{\psi_k, \lambda_k, T_k, Q_k[0], \ldots, Q_k[P-1]\} \).

\textbf{Action (A)}: The action space of the MDP is defined as the set \( A = \{a_1, a_2, \ldots, a_8\} \), where each element corresponds to one of the eight priority signal phases proposed in Section 3.3. Each action \( a_i \) represents the transition of the signal from the current phase to Phase \( i \) state.

\textbf{Probability of State transition (P)}: This function describes the probability of transitioning from one state to another under a particular action. The state transition probability function $P$ records the probability of transitioning from state $s$ to $s'$ after taking action while receiving reward $r$. Denoting the probability symbol by $P$, the state transition probability can be expressed as:

\begin{equation}\label{eq5}
P(s', r \mid s, a) = \mathbb{P}[S_{t+1} = s', R_{t+1} = r \mid S_t = s, A_t = a]
\end{equation}

The state transition function can be defined as:

\begin{equation}\label{eq6}
\begin{aligned}
P_{s,s'}^a &= P(s' \mid s,a) = \mathbb{P}[S_{t+1} = s' \mid S_t = s, A_t = a] \\
&= \sum_{r \in R} P(s', r \mid s, a)
\end{aligned}
\end{equation}

\textbf{Reward function (R)}: The reward function R predicts the next reward triggered by an action.

\begin{equation}\label{eq7}
R(s, a) = \mathbb{E}[R_{t+1} \mid S_t = s, A_t = a] = \sum_{r \in R} r \sum_{s' \in S} P(s', r \mid s, a)
\end{equation}

Specifically, define $i$ as a preferred vehicle intelligence and $j$ as a private vehicle intelligence. In the cooperative game, intelligence $i$ receives the influence of neighbor $j$ in the current state, then the return of $i$ can be defined as a fixed matrix $R_i$ that:

\begin{equation}\label{eq8}
R_i = \begin{bmatrix}
r_i^{11} & \cdots & r_i^{18} \\
\vdots & \ddots & \vdots \\
r_i^{81} & \cdots & r_i^{88} \\
\end{bmatrix}
\end{equation}

where $r_i^{mn}$ denotes the payoff received by intelligent body $i$ when $i$ chooses action $m$ and $j$ chooses action $n$. Iterate over all possible coalitions $T$ , and sum their eigenfunction values to get the expected payoff of the intelligent body $i$ :

\begin{equation}\label{eq9}
E_i(V) = \sum_{T \subseteq V} R(T)
\end{equation}

Where \(V \in 2^{\{i, j\}}\) denotes the coalition formed by intelligences \(i\) and \(j\). \(T \subseteq V\) indicates that the coalition \(T\) is a subset of the coalition \(V\).

The strategy \(\pi\) serves as a representation of the intelligence's behavior, guiding us on the action to take in state \(s\). It is a mapping from state \(s\) to action \(a\) and can be either deterministic or stochastic.A deterministic strategy is denoted as a mapping from \(s\) to \(a\):

\begin{equation}\label{eq10}
\pi(s) = (\pi(1), \pi(2), \ldots, \pi(n)), \quad \pi(s) \in A(s), \quad s = 1,2,\ldots,n
\end{equation}

Probability $p_s^{\pi}$ of adopting strategy $\pi$ in state \(s\):

\begin{equation}\label{eq11}
p_s^\pi = \prod_{s=1}^n p(s_n \mid s_1, \ldots, s_{n-1}) \approx p(s_1) \times \prod_{s=2}^n p(s_n \mid s_{n-1})
\end{equation}

The value function $G$, also known as cumulative reward, is a measure of how good a state is or how rewarding a state or a behavior is by predicting future rewards. Considering rewards at infinity is inappropriate and detrimental to the solution of the MDP, the discounting mechanism introduced in the reward function calculates the cumulative reward by using the sum of future discounted rewards:

\begin{equation}\label{eq12}
G = R_1 + \gamma R_2 + \gamma^2 R_3 + \ldots = \sum_{k=0}^\infty \gamma^k E_{k+1}
\end{equation}

where \(\gamma \in [0,1]\) is a constant, also called the decay factor or discount factor.
Considering the objectives of the TSP system, the reward function quantifies the desirability of each state-action pair. In this case, the reward should promote delay minimization for the priority vehicles while maintaining overall traffic efficiency.
The basic idea of Q-learning is that we define a function Q as follows:

\begin{equation}\label{eq13}
Q^*(s, a) = E(s, a) + \gamma \sum_{s'} p(s' \mid s, a) v(s', \pi^*)
\end{equation}

Where $Q^*(s,a)$ is the total discounted reward for taking action $a$ in state $s$ and following the optimal strategy.

\subsection{Iterative process and solution}\label{sec4.3}

\begin{figure}[!t]
	\centering
	\includegraphics[width=8cm]{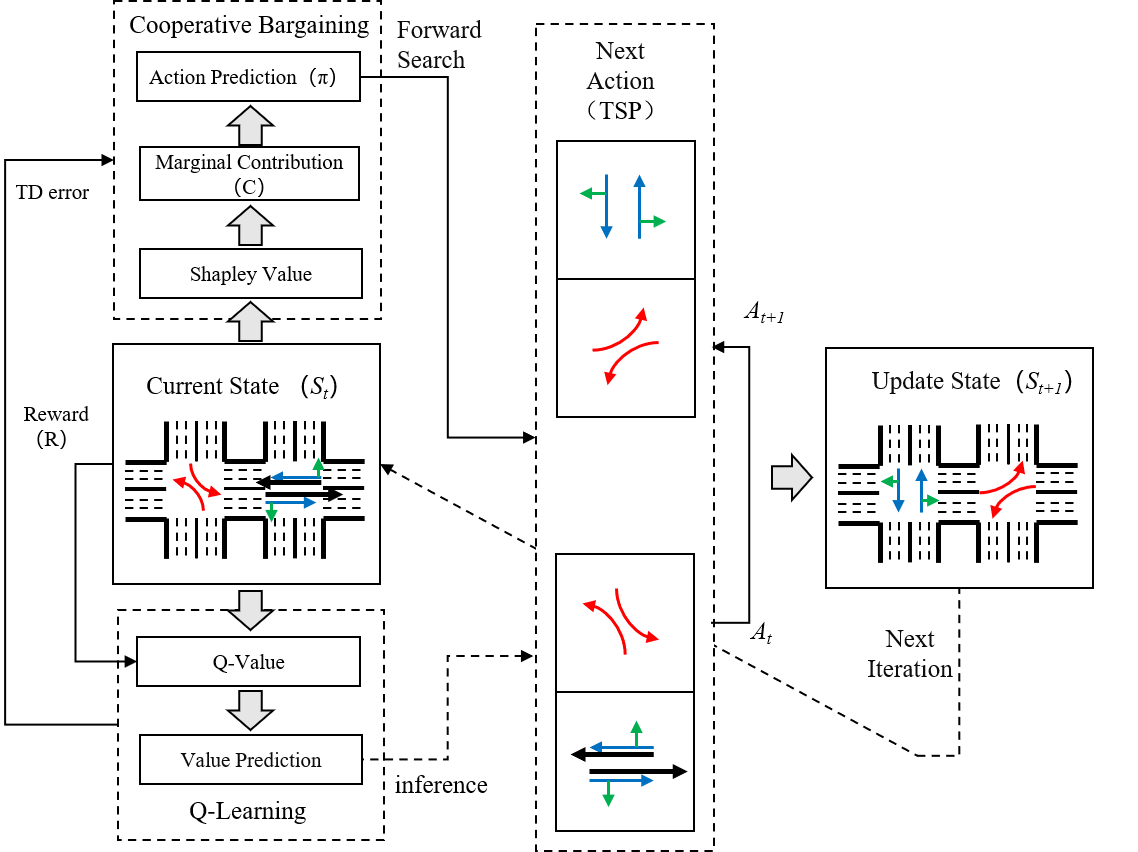}
	\caption{The iterative process between the agents and the environment}
	\label{fig:fig6}
\end{figure}

The basic idea of the CBQL-TSP method is to pass the current state of the traffic environment to two models. The CB model calculates action transition probabilities based on the state, and selects actions based on these probabilities. The execution of an action changes the state of the environment, creating a new state for the next action interval, which is then evaluated for real-time rewards. The real-time rewards and the new state are passed to the QL model. It calculates the Q-values from the action to that particular state, while CB updates its strategy parameters using the Q-values computed by QL. CB calculates the next action based on the updated Q-values and the new state, while the QL model updates its own weights. The iterative process of the CBQL-TSP method is illustrated in Fig.~\ref{fig:fig6}.

Initialize $Q(s,a)$ for all \(s \in S, \quad a \in A\) and update the Q-value as follows:

\begin{equation}\label{eq14}
Q^*(s, a) = E(s, a) + \gamma \sum_{s'} p(s' \mid s, a) v(s', \pi^*)
\end{equation}

Where $a_t$ denotes an estimate of the proportion of all returns that receive the maximum return when action $a$ is chosen.

\begin{equation}\label{eq15}
a_t = \begin{cases}
1 & \text{if } E_t > Q_{\text{max}}(a) \\
(1 - a_f) a_t + a_f & \text{if } E_t = Q_{\text{max}}(a) \\
(1 - a_f) a_t + a_f & \text{if } E_t < Q_{\text{max}}(a)
\end{cases}
\end{equation}

where $a_f$ is the learning rate. For each action $a$ , the algorithm records the maximum payoff $Q_{max} (a)$ that the intelligence has ever received under that action in past experience while computing an updated $Q$. It is worth noting that if it is a single-state environment, the Q-value is an estimate of the immediate payoff $r$.
The joint decision-making strategy is updated in such a way as to gradually increase the probability of selecting the action with the largest Q and decrease the probability of other actions being selected:

\begin{equation}\label{eq16}
\pi(s, a) \leftarrow \pi(s, a) + \begin{cases}
a_f & \text{if } a = \arg\max_{a'} Q(s, a') \\
-\frac{a_f}{|A| - 1} & \text{otherwise}
\end{cases}
\end{equation}

Using the Shapley Value function to calculate the marginal contributions \( C \) of the members in the current game as defined by \( Q \). Denote $I={1,2,\dots,n}$ as the set of $n$ QL Controllers, and define $i $ as one of the members. Denote by $S \backslash \{i\}$ the set $S$ after removing $i$ from the set $S$. Then $v(S \backslash \{i\})$ denotes the revenue of the coalition after $i$ is removed. The marginal contribution is the difference between the coalition's gain and the coalition's gain after removing member $i$. This is the gain contribution that member $i$ brings to the coalition, also called the marginal contribution. So the marginal contribution C of member $i$ in the coalition can be expressed as:
\begin{equation}\label{eq17}
C = v(S) - v(S \backslash \{i\})
\end{equation}

Then $v_{Shapley}^i$ denotes the Shapley value of member $i$, the contribution of member $i$, which can be expressed as:

\begin{equation}\label{eq18}
v_{\text{Shapley}}^i(v) = \sum_{s \in S_i} \omega(\lvert S\rvert) \cdot C
\end{equation}

where \(S_i\) is the set formed by all subsets of \(I\) containing member \(i\), $\lvert S\rvert$ is the number of elements of the set \(S\), \(\omega\) is the weight factor in the coalition, and \(\omega\) is calculated as:

\begin{equation}\label{eq19}
\omega(\lvert S\rvert) = \frac{\lvert S\rvert!(n-\lvert S\rvert-1)!}{n!}
\end{equation}

where \(n\) denotes the total number of collaborators in the coalition.

In the Shapley value calculation process, the study traverses all possible cooperative combinations \(S\) and calculates the new contribution of participant \(i\) after joining the combination. It then performs a weighted average of the new contributions of all combinations, where the weights are determined by the size of the combination.

At time step $t$, agent observes the current state $s$ and takes action, and then observes its own reward $R_t^i$, other agents' actions, other agents' reward $R_t'$, and the new state $s'$. According to eq\eqref{eq14} and eq\eqref{eq18}, the Q value of CBQL can be updated as:

\begin{equation}\label{eq20}
H_{\text{Shapley}}Q(s, a) = (1 - \alpha_t) Q_t^i(s, a) + \alpha_t [R_t^i + \gamma v_{\text{Shapley}}^i(v)]
\end{equation}

\section{Traffic Flow Simulation Platform Construction and Simulation Test}\label{sec5}
\begin{figure}[!t]
	\centering
	\includegraphics[width=10cm]{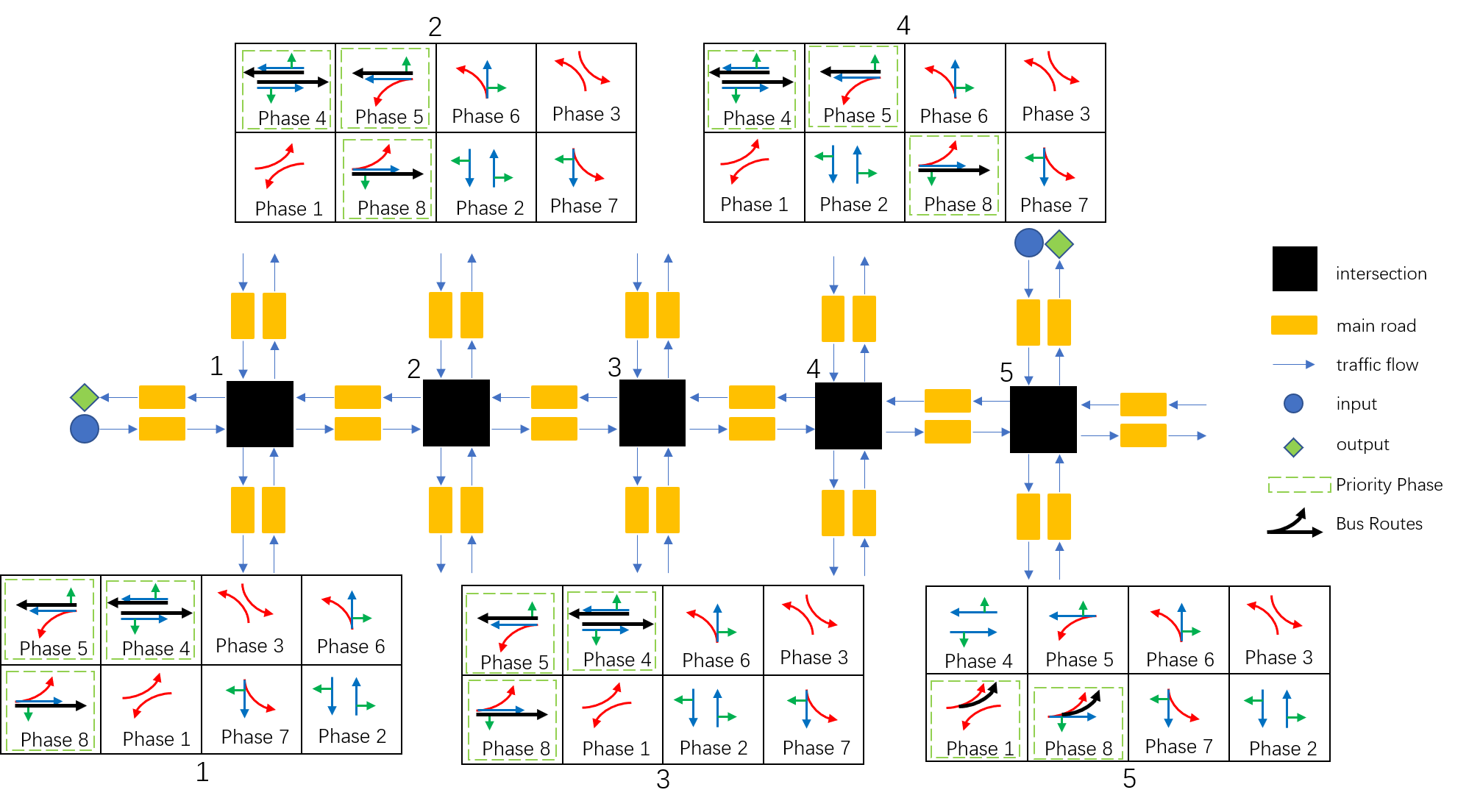}
	\caption{Test model with five coordinated intersections}
	\label{fig:fig7}
\end{figure}

To evaluate the effectiveness of the joint control strategy proposed in the paper, experimental validation was conducted, targeting application in urban regional traffic environments. The experiments utilized the microscopic traffic simulation software PARAMICS, which has a Python interface, to create a road network model containing a dedicated bus lane at intersections for defining the study area. As shown in Fig.~\ref{fig:fig7}, Dalian High-tech Zone has a dedicated lane for priority vehicles with five intersections. Through investigation, it was determined that the traffic volume on this road is sufficiently high during peak hours, making it suitable for experimentation.

Initially, lane data including speed limits, traffic volumes, and counts was converted to OpenStreetMap format and imported into PARAMICS. Traffic flow data from Line 10 buses at five Dalian intersections was processed to align with PARAMICS' specifications for vehicle positions, speeds, and directions. Signal control settings for each intersection were established, including phase numbers and durations. The network model was validated, and an eight-phase signal control system was implemented in PARAMICS to test the priority signal control.

In the experiment, three recently effective TSP control methods were chosen as comparative algorithms. These algorithms, proposed in recent years, have demonstrated favorable performance in this field. The first one is Microscopic Simulation-based TSP (referred to as MB-TSP)\cite{bib25} , which falls under passive TSP. The second is Active TSP based on Maximum Pressure Control Algorithm (referred to as MP-TSP)\cite{bib20} . The third is Adaptive Signal Control with consideration for Bus Signal Priority (referred to as ASC-TSP), belonging to adaptive TSP\cite{bib23}. It is worth noting that bus signal priority is only granted to dedicated bus lanes in this study. In instances where there are no dedicated bus lanes on BRT routes, any signal controller in this simulation will be unable to implement bus signal priority. In this simulation, there is no conflicting movement in the dedicated bus lanes within the network. For all traffic signal control strategies in this paper, the bus signal priority strategy remains consistent.

\subsection{Stability Comparison}\label{sec5.1}
First, for the definition of stability: a network is considered stable if the number of private cars in the network maintains a bounded expectation. That is, there exists an $s < \infty$ such that.

\begin{equation}\label{eq21}
\lim_{{T \to \infty}} \sup \left\{ \frac{1}{T} \sum_{{t=1}}^{T} \sum_{{(i,j) \in A^2}} E\{x_{ij}^P(t)\} \leq s \right\}
\end{equation}

When private car demand is within the stable region, the average private car ownership will converge to a constant. However, for unstable demand, the average number of private vehicles will increase to an arbitrarily large number.

\begin{figure}[!t]
	\centering
	\includegraphics[width=8cm]{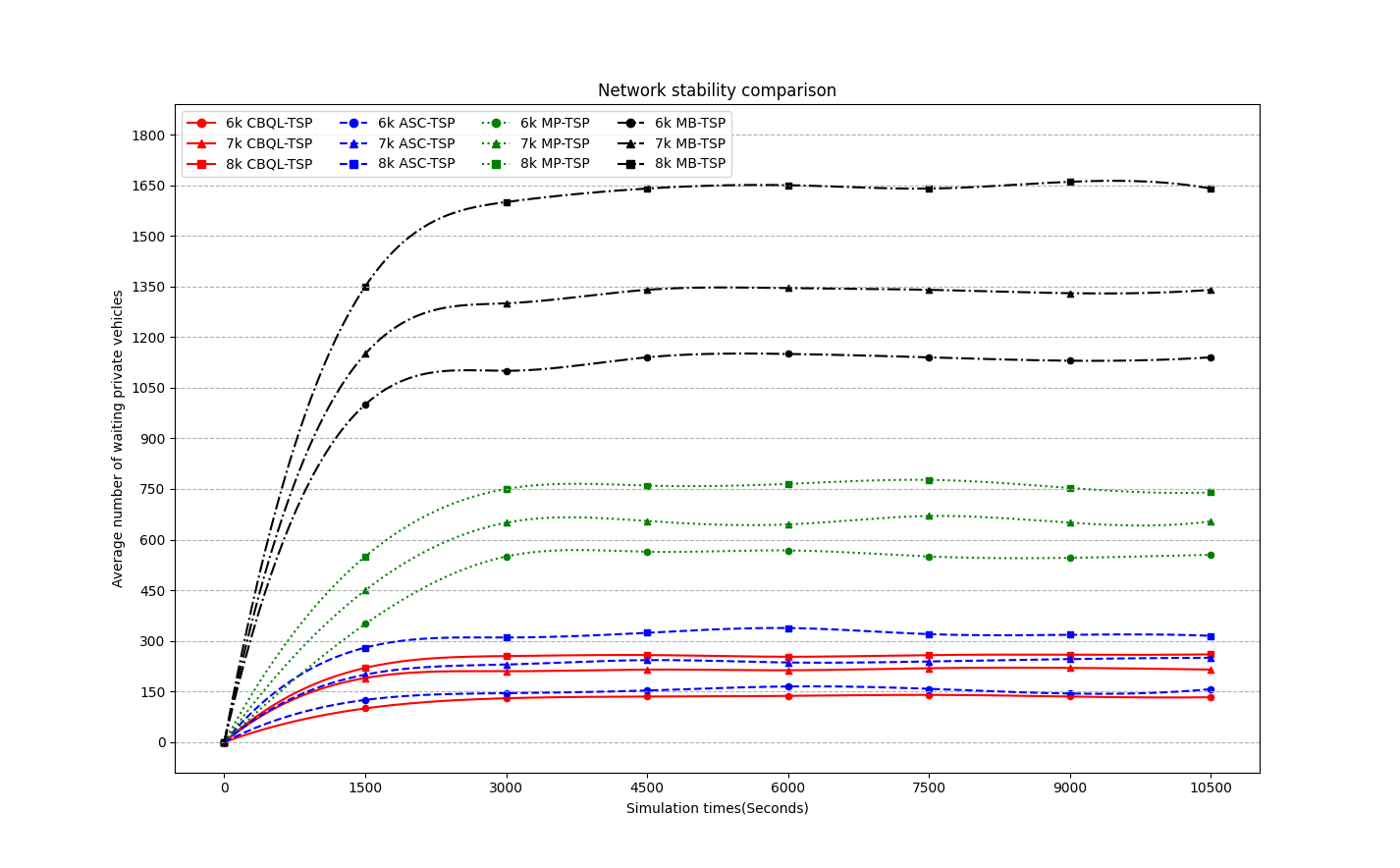}
	\caption{Test result of Network stability comparison}
	\label{fig:fig8}
\end{figure}

Fig.~\ref{fig:fig8} compares the results of the average waiting number of private cars for MP-TSP, MB-TSP, ASC-TSP, and CBQL-TSP. Under the same private car demand settings, CBQL-TSP consistently exhibits lower private car waiting numbers compared to the experimental results of MP-TSP, MB-TSP, and ASC-TSP. Furthermore, as the total number of private cars increases from 6k to 8k, the average waiting number of private cars for all four methods increases. In comparison to MB-TSP and ASC-TSP, MP-TSP and CBQL-TSP demonstrate better stability regions, with CBQL-TSP having a lower waiting number of private cars. These results indicate that CBQL-TSP has a larger stability region compared to MP-TSP, MB-TSP, and ASC-TSP, and it can maintain network stability at higher levels of private car demand.

\begin{figure}[!t]
	\centering
	\includegraphics[width=8cm]{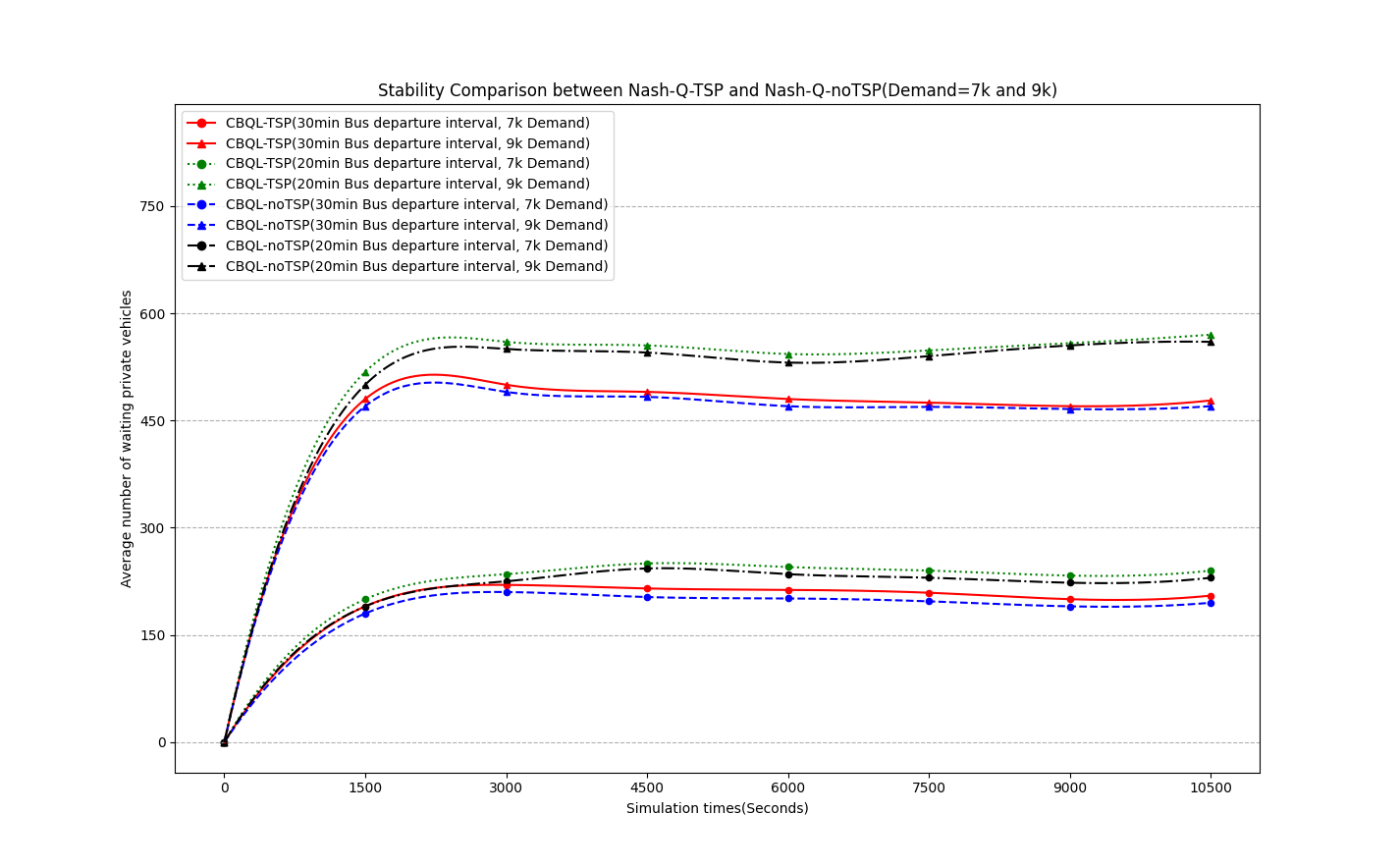}
	\caption{Test result of stability comparison between CBQL-TSP and CBQL-noTSP}
	\label{fig:fig9}
\end{figure}

We investigated the impact of signal priority strategies on intersection stability, analyzing private car wait times under varying demands and control methods. Comparing CBQL signal control without transit signal priority (CBQL-noTSP) and an eight-phase transit signal priority (CBQL-TSP) under different bus departure intervals (20 and 30 minutes), our findings are illustrated in Fig.~\ref{fig:fig9}. Under the same demand settings, CBQL-noTSP resulted in fewer waiting private cars, as bus priority reduces private car right of way. However, with shorter bus departure intervals, private car wait times increased under CBQL-TSP due to more time allocated to buses. When demand increased from 7k to 9k, both strategies saw about 300 more waiting cars. Switching from CBQL-noTSP to CBQL-TSP added approximately 10 more waiting cars under various demands, showing the algorithm's minimal and stable impact on private car wait times across different traffic demands.

\subsection{Analysis of Travelling Time Test Results}\label{sec5.2}

\begin{figure}[!t]
	\centering
	\includegraphics[width=8cm]{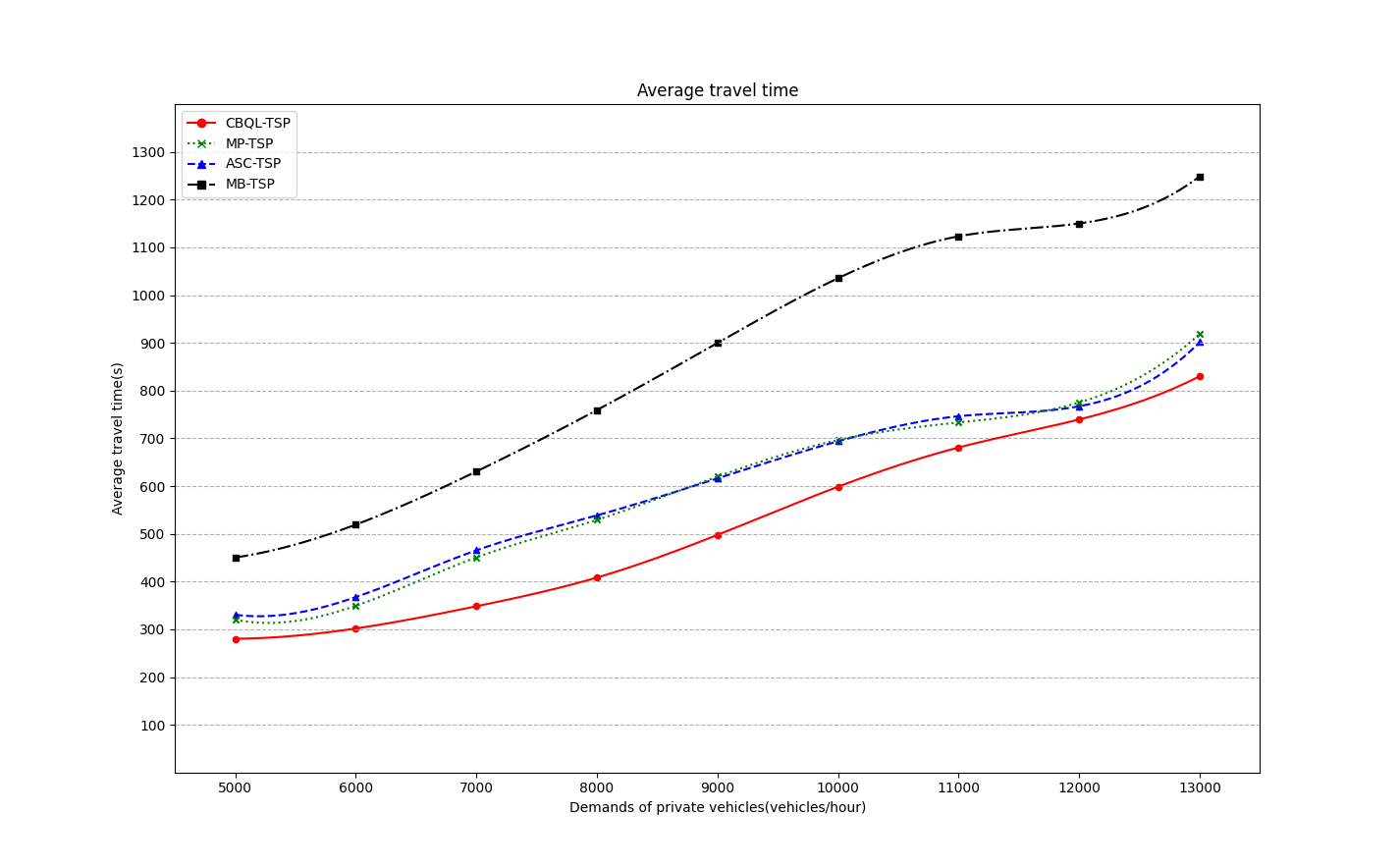}
	\caption{Test result of Average travel time}
	\label{fig:fig10}
\end{figure}

In addition to the stability comparison, it is also necessary to explore how bus signal prioritization affects vehicle travel times at the network level. The average bus travel times for CBQL-TSP, ASC-TSP, MP-TSP, and MB-TSP are shown in Fig.~\ref{fig:fig10} for a 30 min bus departure interval. As the demand for private vehicles continues to increase, vehicles spend more time on roadway segments and intersections.

As can be seen in Fig.~\ref{fig:fig10}, MB-TSP has the most average travel time, and the effects of MP-TSP and ASC-TSP each have their own advantages in terms of performance for different private vehicle demands, and the algorithm proposed in this paper has a shorter average bus travel time compared to the other three algorithms for any demand, and the effect is more obvious especially when the demand for private vehicles is between 7k and 10k. For better comparison, the study summarizes the experimental results in Table~\ref{tab:tab1}.
The average transit times for MB-TSP, MP-TSP, and ASC-TSP are 868.53 seconds, 599.19 seconds, and 603.00 seconds, respectively. The average transit time for CBQL-TSP is approximately 520.63 seconds, representing a reduction of about 24.57\% compared to the other algorithms.

\begin{table}
	\centering
	\caption{Experimental results of average travel time }
	\label{tab:tab1}
	\begin{tabular}{|c|c|c|c|c|}
		\hline
		Demands & CBQL-TSP & ASC-TSP & MP-TSP  & MB-TSP \\
		(vehicles/h) &  (s) &  (s) & (s) &  (s)\\
		\hline
		5000 & 280.21 & 330.47 & 320.21 & 450.41 \\
		6000 & 301.72 & 367.51 & 349.31 & 519.31 \\
		7000 & 348.24 & 465.49 & 450.58 & 630.58 \\
		8000 & 408.41 & 538.55 & 529.18 & 759.18 \\
		9000 & 497.76 & 616.15 & 619.82 & 899.82 \\
		10000 & 598.68 & 693.89 & 696.44 & 1035.44 \\
		11000 & 680.84 & 746.35 & 733.21 & 1123.21 \\
		12000 & 739.57 & 767.17 & 775.10 & 1150.10 \\
		13000 & 830.21 & 901.41 & 918.82 & 1248.75 \\
		\hline
	\end{tabular}
\end{table}

\begin{figure}
	\centering
	\includegraphics[width=8cm]{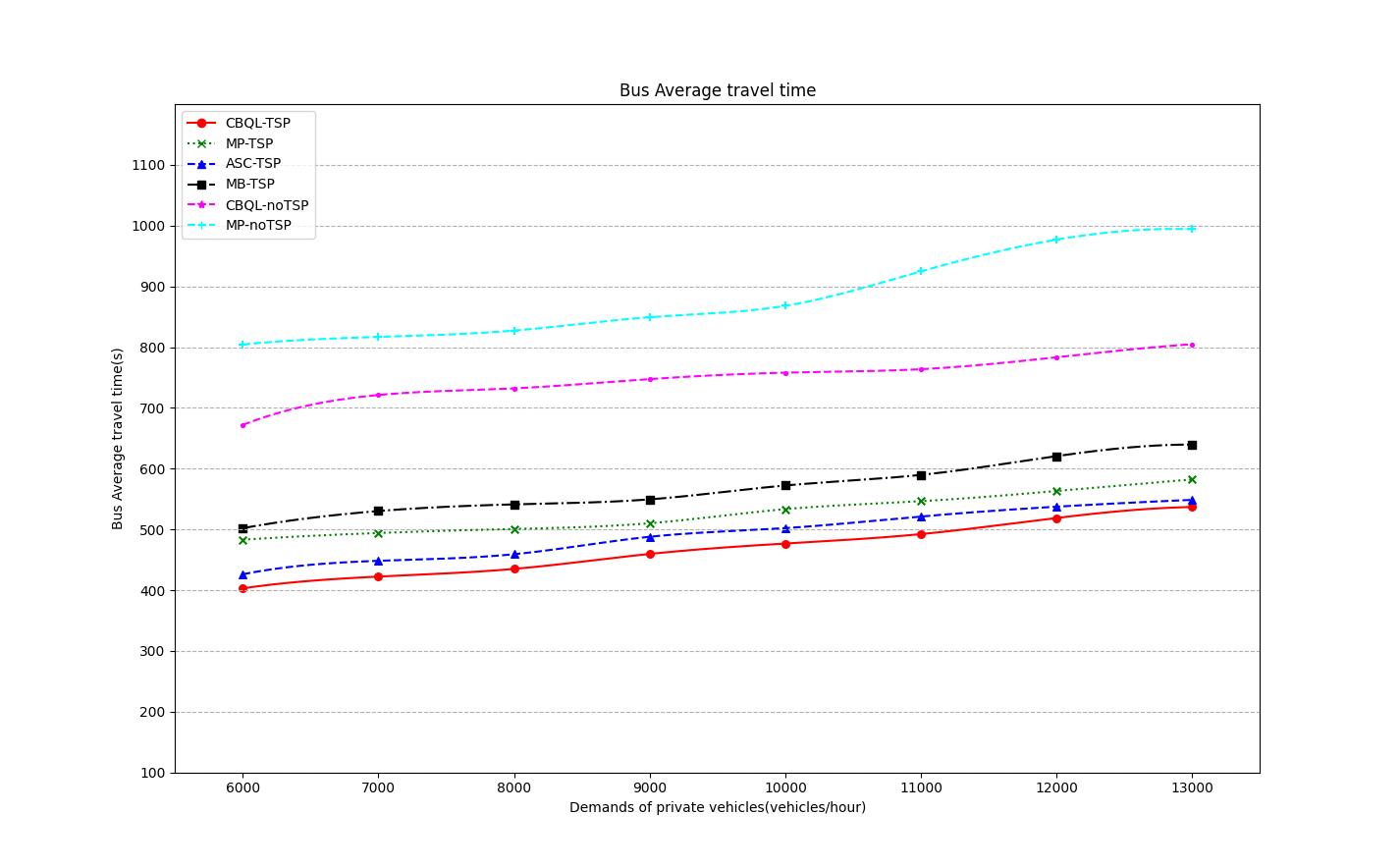}
	\caption{Test result of Average travel time}
	\label{fig:fig11}
\end{figure}

Next, bus travel time fluctuations on test arterials were tested after the implementation of joint decision control and transit signal prioritization. The study compares the average travel times of buses through five intersections for four algorithms, Maximum Pressure Control without Signal Priority (referred to as MP-noTSP)\cite{bib26}, ASC-TSP, MP-TSP, MB-TSP, CBQL-noTSP, and CBQL-TSP, for a fixed period of time. The results are shown in Fig.~\ref{fig:fig11}. As the demand for private cars increases, the average travel time of buses on the test arterials increases under all five traffic signal controllers. The average travel time of buses is the highest under the MP-noTSP controller. The CBQL-noTSP controller reduces the average travel time of the buses in the central city, but it is not optimal.

In addition, the average bus travel times in the downtown area without bus signal priority (FT-noTSP and CBQL-noTSP) are all greater than the bus travel times under the traffic signal controllers with bus signal priority (MB-TSP, ASC-TSP and CBQL-TSP). ASC-TSP is based on more loop detectors than MB-TSP, FT-noTSP and MP-noTSP, so it is the second most effective. Finally, it was found that CBQL-TSP reduces the bus travel time in the downtown area when the interval between two bus departures is set to 30 minutes. Table~\ref{tab:tab2} shows the specific data of the experimental results.

The average transit time for buses in the city center is approximately 468.22 seconds with the CBQL-TSP method, compared to 747.93 seconds with the CBQL-noTSP method, indicating that the TSP method reduces the average bus transit time in the city center by about 37.40\%. Meanwhile, the average transit times for buses in the city center with MB-TSP, ASC-TSP, and another iteration of MB-TSP are 491.50 seconds, 526.79 seconds, and 568.28 seconds, respectively. Compared to these three TSP methods, the CBQL-TSP method reduces the average bus transit time in the city center by approximately 11.46\%.

\begin{table}[!t]
	\centering
	\caption{Results of bus average travel time}
	\label{tab:tab2}
	\resizebox{\linewidth}{!}{
		\begin{tabular}{|c|c|c|c|c|c|c|}
			\hline
			Demands  & CBQL- & MP-TSP & ASC-TSP & MB-TSP & CBQL-& MP-noTSP \\
			(vehicles/h) &  TSP(s) &  (s) &  (s) &  (s) &  noTSP(s) &  (s)\\
			\hline
			6000 & 403.12 & 426.32 & 483.26 & 502.21 & 672.12 & 804.51 \\
			7000 & 422.45 & 448.44 & 494.24 & 530.34 & 721.33 & 816.91 \\
			8000 & 435.12 & 459.31 & 500.85 & 541.32 & 732.27 & 827.38 \\
			9000 & 459.76 & 488.01 & 510.26 & 549.58 & 747.49 & 849.33 \\
			10000 & 476.88 & 502.27 & 533.50 & 572.42 & 758.09 & 868.12 \\
			11000 & 492.40 & 521.35 & 546.57 & 589.85 & 763.79 & 924.82 \\
			12000 & 518.96 & 537.52 & 563.37 & 620.74 & 783.44 & 977.03 \\
			13000 & 537.03 & 548.75 & 582.25 & 639.77 & 804.94 & 994.20 \\
			\hline
		\end{tabular}
	}
\end{table}

\section{Conclusions}\label{sec6}

This paper proposes an eight-phase priority signal control method, dividing the signal control sequence into priority and non-priority signals for study. A hybrid decision model is constructed to explore the cooperative game between priority and non-priority signals. By solving the Shapley value function, the Shapley value ratios of the cooperative game members are obtained, and these ratios determine the MDP state transition probabilities. Comparative experiments were designed, and the results demonstrate that the proposed method offers better stability at intersections with bus priority, reducing the overall average transit time by approximately 24.57\% and the bus average transit time by about 37.40\% compared to other TSP methods and no TSP method, respectively. This research implements a bus signal priority strategy based on cooperative games and reinforcement learning, balancing vehicle priority from the perspective of marginal contributions. By reasonably allocating intersection priorities, the method enhances the transit efficiency of public vehicles at intersections while also ensuring the transit efficiency of private vehicles, thus improving the overall operation efficiency of the intersection. The proposed method not only enhances the overall performance of the traffic system but also boosts the sustainability and fairness of urban traffic, creating a more balanced and efficient driving environment for all road users. For future research directions, introducing a multi-level priority signal control system, especially regarding the internal stratification of priorities, remains an area for further solution development. Consideration could be given to incorporating machine learning algorithms to provide fine-grained control over priorities based on the traffic behavior patterns and impacts of priority vehicles, further optimizing priority signal control methods.
\section{Acknowledgments}
We would like to thank the anonymous reviewers for providing constructive feedback and precious comments.


\end{document}